\documentclass[a4paper,11pt]{article}
\pdfoutput=1 

\usepackage{jheppub} 

\usepackage[T1]{fontenc} 
\usepackage{graphicx}
\usepackage{amsthm}
\usepackage{amsmath}
\usepackage{graphicx,subfigure}
\usepackage{dcolumn}
\usepackage{bm}
\usepackage{slashed}
\usepackage{calligra}

\usepackage{booktabs} 

\title{\boldmath Novel dependence between neutrino mass splittings strongly supported by initial JUNO results}

 \author{I. Alikhanov}
 \affiliation{North-Caucasus Center for Mathematical Research, North-Caucasus Federal University,\\
Stavropol 355017, Russian Federation}

\emailAdd{ialspbu@gmail.com}

\abstract{
In the standard three-flavor paradigm, two of the three neutrino mass-squared differences are independent. The existence of an additional dependence between these oscillation parameters was recently predicted on empirical grounds, taking the form
$$\frac{\sqrt{\Delta m^2_{31}}+\sqrt{\Delta m^2_{21}}}{\sqrt{\Delta m^2_{31}}-\sqrt{\Delta m^2_{21}}}=\sqrt{2}.$$
The JUNO Collaboration has now reported the most precise measurements to date of both~$\Delta m^2_{21}$ and $\Delta m^2_{31}$. We demonstrate that this relation is in excellent agreement with the experimental data, which yield $1.4143_{-0.0038}^{+0.0036}$ corresponding to a relative precision of~$0.27\%$. Remarkably, the central value differs by less than $0.03$ standard deviations from the prediction~($\sqrt{2}\approx1.4142$). This provides strong experimental support for the novel constraint and thereby suggests an underlying symmetry governing the neutrino mass spectrum, rather than purely independent parameters.
}

\begin{document} 
\maketitle
\flushbottom

\section{Introduction}
In the standard three-flavor paradigm, three neutrino flavor eigenstates ($\nu_e$,~$\nu_\mu$,~$\nu_\tau$) are mixed with three mass eigenstates ($\nu_1$, $\nu_2$, $\nu_3$) of masses $m_1$, $m_2$, $m_3$. This mixing is described by the unitary $3\times3$ Pontecorvo--Maki--Nakagawa--Sakata matrix~\cite{ParticleDataGroup:2024cfk}. 
The validity of this framework has been confirmed in a relatively wide range of neutrino energies and for different neutrino sources, both natural and artificial, such as the Sun~\cite{SNO:2002tuh,Super-Kamiokande:2001ljr}, the atmosphere~\cite{Super-Kamiokande:1998kpq,IceCube:2014flw}, accelerators~\cite{K2K:2004iot,MINOS:2008kxu,T2K:2011ypd,OPERA:2015wbl}, and nuclear reactors~\cite{KamLAND:2002uet,DoubleChooz:2011ymz,DayaBay:2012fng,RENO:2012mkc}.  Since the standard oscillation mechanism requires nonzero masses, the results of these experiments also demonstrate that at least two of the three neutrino mass eigenstates are massive. However, oscillation experiments are sensitive only to the neutrino mass-squared differences rather than the absolute masses themselves.
Determining the latter remains a fundamental yet unsolved problem in particle physics. The Standard Model does not predict the exact neutrino masses. 

Two orderings of neutrino masses are possible: i)~normal, in which~$m_1<m_2<m_3$, and ii)~inverted ordering, which implies~$m_3<m_1<m_2$. Current data favor the first possibility~\cite{T2K:2023smv,T2K:2025yoy,NOvA:2025tmb, Esteban:2026phq, JUNO:2026tap}. The operating JUNO detector~\cite{JUNO:2022mxj,JUNO:2025gmd}, along with upcoming projects such as DUNE~\cite{DUNE:2020lwj} and Hyper-Kamiokande~\cite{Hyper-Kamiokande:2018ofw}, aims to determine the true mass ordering with high statistical significance ($3\text{--}5\sigma$).
In what follows, we assume normal mass ordering.

There are three mass-squared differences, $\Delta m^2_{21}=m^2_2-m^2_1$, $\Delta m^2_{31}=m^2_3-m^2_1$, and \mbox{$\Delta m^2_{32}=m^2_3-m^2_2$}, of which only two are  independent in the standard three-flavor oscillation paradigm~\cite{Denton:2025jkt}. They are linked by their  consistency condition: $\Delta m^2_{31}-\Delta m^2_{21}=\Delta m^2_{32}$. However, the existence of an additional simple constraint on these parameters has recently been predicted on phenomenological grounds~\cite{Alikhanov:2026jmw}. Traditionally, $\Delta m^2_{21}$ and $\Delta m^2_{31}$  are also referred to as the solar and atmospheric neutrino mass splittings, respectively. The JUNO Collaboration has just reported the most precise measurements to date of both splittings~\cite{JUNO:2026tap}~(see Table~\ref{tab:tab1}). These parameters will be pinned down at a sub-percent level in the near future as the experiment continues to accumulate data~\cite{JUNO:2022mxj}. Its  extraordinary precision offers a pristine environment to test the robustness of the three-flavor paradigm. 
\begin{table}[h!]
 \begin{center}
      \caption{The solar and atmospheric neutrino mass splittings measured by JUNO. The results correspond to $207.2$~live days of data assuming normal mass ordering~\cite{JUNO:2026tap}.}
      \medskip
\setlength\extrarowheight{3pt}
      \begin{tabular}{|c c c|}\hline 
      
      {Parameter}\rule{25pt}{0pt}&{ Central value}\rule{25pt}{0pt} & { $1\sigma$ range} \\[1em]
       \hline 
  \rule{0pt}{20pt}   {$\dfrac{\Delta m^2_{21}}{10^{-5}~\text{eV}^2}$}\rule{25pt}{0pt}  &$7.388$\rule{25pt}{0pt} &$7.310\rightarrow7.466$ \\[1em]
    \hline\rule{0pt}{20pt}
     {$\dfrac{\Delta m^2_{31}}{10^{-3}~\text{eV}^2}$}\rule{25pt}{0pt} & $2.509$\rule{25pt}{0pt}&$2.484\rightarrow 2.536$\\[1em]
     \hline
           \end{tabular}
      \label{tab:tab1}
  \end{center}
\end{table}
Any statistically significant deviation from the consistency condition would serve as an unambiguous signature of physics beyond the Standard Model, such as exotic neutrino decays, decoherence, or the presence of light sterile states. 

In this work, we demonstrate that the initial JUNO results~\cite{JUNO:2026tap} strongly support the existence of the aforementioned novel dependence between the neutrino mass splittings—beyond their standard consistency condition. We also outline how this relation is expected to crystallize as higher-precision experimental data emerge.  

\section{Comparison with experiment}
Motivated by an analysis of global neutrino data fits~\citep{Capozzi:2017ipn,Capozzi:2025wyn,Esteban:2024eli,Esteban:2026phq}, we have recently argued that~\cite{Alikhanov:2026jmw} 

\begin{equation}
\label{eqref:main}
\dfrac{\sqrt{\Delta m^2_{31}}+\sqrt{\Delta m^2_{21}}}{\sqrt{\Delta m^2_{31}}-\sqrt{\Delta m^2_{21}}}=\sqrt{2}.
\end{equation}
This relation constitutes a novel constraint on the oscillation parameters. If~\eqref{eqref:main}~holds, it reduces the number of independent neutrino mass splittings to just one.  It should be emphasized that the current consensus dictates that two of the three neutrino mass-squared differences are independent~\cite{Denton:2025jkt}. In this framework, the validity of~\eqref{eqref:main}~would imply an underlying mechanism or symmetry governing the neutrino mass spectrum that goes beyond the standard assumptions of the three-flavor paradigm. 
\begin{figure}[htbp]
    \centering
    \includegraphics[width=0.9\textwidth]{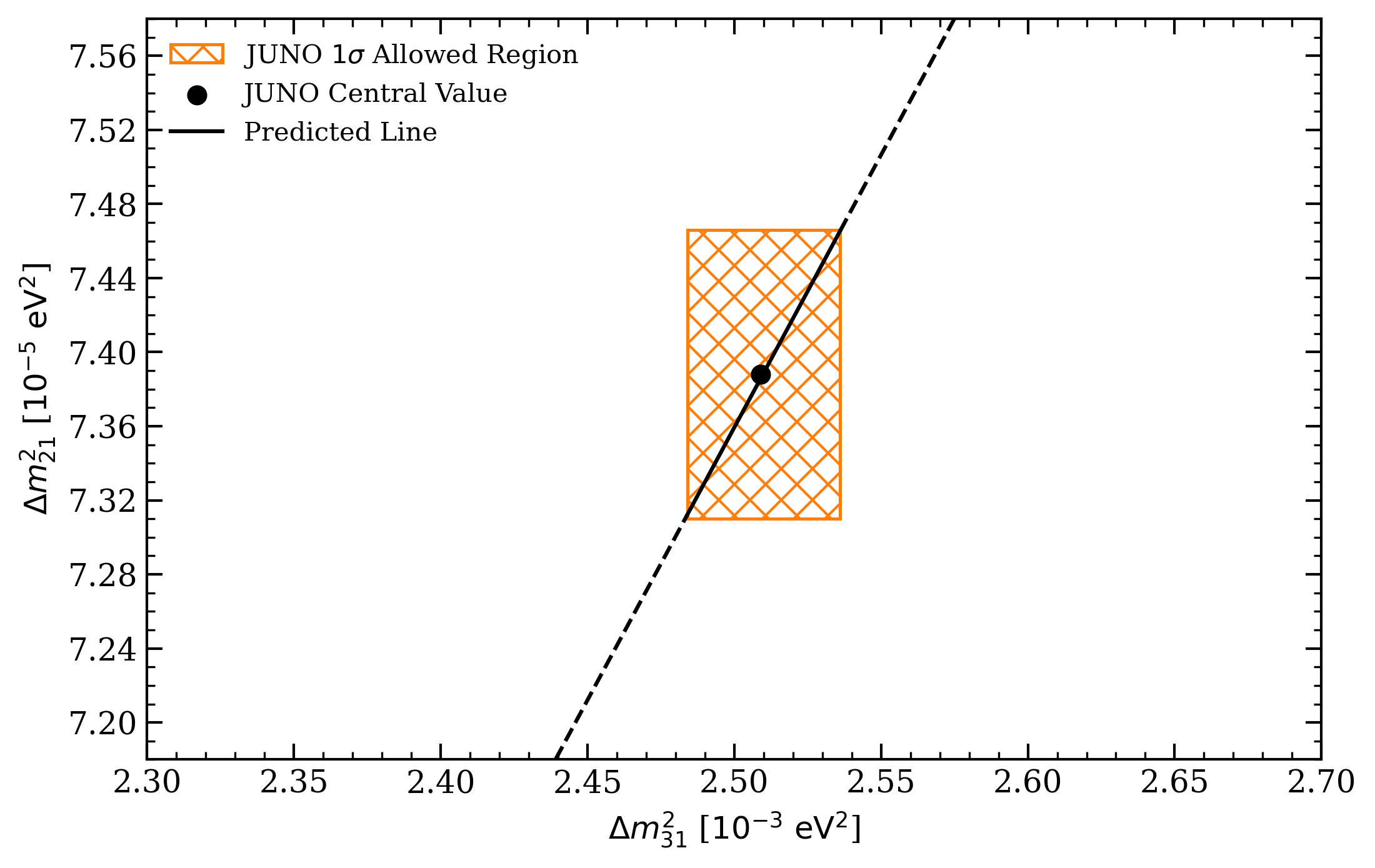}
    \caption{The $(\Delta m^2_{21}, \Delta m^2_{31})$ parameter space. The cross-hatched rectangle represents the $1\sigma$~allowed region from JUNO~\cite{JUNO:2026tap}, which is also given numerically in Table~\ref{tab:tab1}. The solid black point corresponds to the experimental central values. The straight line manifests the existence of the novel dependence, which reduces the two-dimensional parameter space to a single linear trajectory upon which the empirical data points must fall.}
    \label{fig:neutrino}
\end{figure}

The JUNO Collaboration has now reported the most precise measurements to date of both $\Delta m^2_{21}$ and $\Delta m^2_{31}$~\cite{JUNO:2026tap} (see Table~\ref{tab:tab1}).  These unprecedented experimental data allow for a direct and stringent test of the prediction.
Substituting the corresponding values from Table~\ref{tab:tab1} into the left-hand side of~\eqref{eqref:main} and employing the standard first-order error propagation for independent variables, we obtain 
\begin{equation} \label{eq:main_predict22} 
\frac{\sqrt{\Delta m^2_{31}}+\sqrt{\Delta m^2_{21}}}{\sqrt{\Delta m^2_{31}}-\sqrt{\Delta m^2_{21}}} = 1.4143_{-0.0038}^{+0.0036}.
\end{equation}
This result is in excellent agreement with~\eqref{eqref:main}; its central value differs by less than $0.03$~standard deviations from the prediction ($\sqrt{2}\approx1.4142$). Crucially, this exceptional match is achieved at a sub-percent relative experimental precision of~$0.27\%$. Nevertheless, one should bear in mind that the observed level of agreement is preliminary since JUNO continues to accumulate statistics and refine its results. This prospect further enhances the interest in upcoming, higher-precision oscillation data. 

Since the spectral modulations corresponding to the solar and atmospheric splittings are physically distinct within the JUNO reactor antineutrino spectrum, the correlation between these parameters is expected to be negligibly small. This justifies the application of the standard error propagation procedure for independent variables.

Comparing~\eqref{eq:main_predict22} with~\eqref{eqref:main}, it is intuitively difficult to dismiss this match as purely accidental, particularly in light of the appealing symmetry of~\eqref{eqref:main}. Indeed, in particle physics, ratios of observable quantities—such as scattering cross-sections and magnetic moments—as well as their underlying amplitudes, frequently yield clean fractions or simple square roots such as~$\sqrt{2}$, $\sqrt{3}$, or~$\sqrt{6}$~\cite{halzen1984quarks}. A plausible explanation for the emergence of this specific value could therefore be an overall normalization factor or an effect stemming from mixing angles. In any case, deriving this relation within a quantum theoretical framework would provide a rigorous foundation for the phenomenological approach adopted here.
\newpage
\section{Reduction of parameter space}


In this section, we demonstrate that a strict linear relation between the solar and atmospheric neutrino mass splittings directly follows from~\eqref{eqref:main}. Consequently, the allowed parameter space reduces to a straight line. Specifically, solving~\eqref{eqref:main} for~$\Delta m^2_{21}$ yields 
\begin{equation}
\label{eq:solution_m21}
\Delta m^2_{21}=\left(\sqrt{9}-\sqrt{8}\right)^2\Delta m^2_{31},
\end{equation}
which explicitly determines the aforementioned relation. 

It turns out that, under the validity of~\eqref{eqref:main}, the allowed $(\Delta m^2_{21}, \Delta m^2_{31})$ parameter space becomes effectively one-dimensional, shrinking to the trajectory defined by~\eqref{eq:solution_m21}. This, in turn, imposes a stringent requirement on the distribution of experimental data points within the $(\Delta m^2_{21}, \Delta m^2_{31})$ plane: they are theoretically constrained to lie on this line. Figure~\ref{fig:neutrino} displays the initial JUNO result alongside the plot of~\eqref{eq:solution_m21}. Remarkably, the point corresponding to the experimental central values coincides almost perfectly with the line, in excellent agreement with the prediction of the hypothesis. We look forward to refined JUNO data to monitor the evolution of this point with increased exposure; within the proposed framework, any future shift of this point is strictly restricted to this linear trajectory.

\section{Conclusions \label{sec:conclusion}}
In the standard three-flavor paradigm, two of the three neutrino mass-squared differences are customarily treated as independent parameters. In this work, we have demonstrated that the initial JUNO measurements~\cite{JUNO:2026tap} strongly support the existence of a simple yet novel constraint on these oscillation parameters that reduces the number of independent neutrino mass splittings to just one. This constraint was recently predicted on phenomenological grounds in~\citep{Alikhanov:2026jmw} and takes the form of a ratio equal to~$\sqrt{2}$. The accuracy with which the current JUNO data reproduce this ratio is remarkable, yielding $1.4143_{-0.0038}^{+0.0036}$. The experimental central value differs by less than $0.03$ standard deviations from the prediction~($\sqrt{2}\approx1.4142$), thereby suggesting an underlying symmetry governing the neutrino mass spectrum rather than purely independent parameters. Crucially, this exceptional match is achieved at a sub-percent relative experimental precision of~$0.27\%$.

If this additional dependence exists, it reduces the two-dimensional parameter space in the $(\Delta m^2_{21}, \Delta m^2_{31})$ plane to a single linear trajectory. This provides a distinct, testable signature of the constraint: experimental points must fall on this line. We have derived the equation of this linear trajectory and showed that the point corresponding to the central values reported by JUNO lies almost perfectly on it. It remains to be seen how future data  will shift this point.

It is intuitively difficult to dismiss as purely accidental the fact that this famous irrational number can be approximated so accurately by such a simple combination of oscillation parameters. The symmetry of the relation—where the numerator and denominator share the same mathematical structure—and the appearance of this specific numerical value  point toward a cancellation process of the underlying quantum mechanical quantities, yielding an overall normalization factor or an effect stemming from the mixing angles. Indeed, numerous textbook examples from particle physics feature ratios of amplitudes that yield simple square roots such as~$\sqrt{2}$, $\sqrt{3}$, or~$\sqrt{6}$.  In any case, deriving this relation within a quantum-field-theoretical framework would provide a rigorous foundation for this empirical result.

Throughout this work, we have assumed the normal mass ordering scenario. Nevertheless, it is instructive to examine how the relation behaves under the inverted ordering hypothesis. Substituting the absolute value of the corresponding atmospheric splitting  measured by JUNO~\cite{JUNO:2026tap} into the ratio yields $1.4170 \pm 0.0037$. Although this result agrees with the $\sqrt{2}$ prediction within the $1\sigma$ interval, its central value falls visibly farther from it, exhibiting a $0.75\sigma$ deviation. In this sense, the relation appears to favor the normal ordering scenario, at least preliminarily. It will be interesting to verify with upcoming data whether this prospective sensitivity to the mass ordering scenarios persists.

The current JUNO data are still at a relatively early stage, being based on~$207.2$ live days, yet the experiment has already achieved a world-leading precision of approximately~$1\%$ for the neutrino mass splittings~\cite{JUNO:2026tap}. As the measurements are ongoing and more refined experimental results are anticipated, we look forward to seeing whether they will sustain these intriguing  initial findings.


\acknowledgments
The author thanks J.~F.~Beacom for useful correspondence.


\bibliographystyle{JHEP}
\bibliography{refs}
\end{document}